\documentstyle[epsf,12pt]{article}
\newcommand{\z}{&\hspace*{-8pt}}

\begin{document}

\begin{flushright}
BI-TP-97/43 \\
\end{flushright}

\begin{center}

\vspace*{40mm}

{\Large \bf Large Mass Expansion versus Small Momentum Expansion of
Feynman Diagrams}

\vskip 10mm

J.~Fleischer
\footnote{~E-mail: fleischer@physik.uni-bielefeld.de}
M.~Yu.~Kalmykov
\footnote{Bogoliubov Laboratory of Theoretical Physics,
Joint Institute for Nuclear Research, 141980, Dubna (Moscow Region), Russia}
\footnote{~E-mail:kalmykov@thsun1.jinr.dubna.su}
\footnote{~Supported by Volkswagen-Stiftung under I/71~293 }
and
O.~L.~Veretin
\footnote{~E-mail:veretin@physik.uni-bielefeld.de}
\footnote{~Supported by BMBF under 05~7BI92P(9)}

\vskip 10mm

{\it ~Fakult\"at f\"ur Physik, Universit\"at Bielefeld,
D-33615 Bielefeld, Germany.}

\begin{abstract}

The method of the large mass expansion (LME) has the technical advantage 
that two-loop integrals occur only as bubbles 
with large masses. In many cases only one large mass occurs.
In such cases these integrals
are expressible in terms of $\Gamma$-functions, i.e. they can be handled
completely analytically avoiding even recursions and therefore 
this approach may find a wide field
of application. We consider it necessary to investigate the precision of
this method and test it for several two-loop
vertex functions ocurring in the $Z \to b\bar{b}$ decay
by comparing it with the small momentum expansion.
It turns out that in general high order approximants have to be taken into
account for a sufficient accuracy.

\end{abstract}
\end{center}

{\it PACS numbers}: 12.15.Ji; 12.15.Lk; 13.40.-b; 12.38.Bx; 11.10.Ji
\\

{\it Keywords}: Standard Model; Feynman diagram;
two-loop diagram; vertex diagram.
\vfill

\thispagestyle{empty}
\setcounter{page}0
\newpage

\section{Introduction}

The experiments at LEP/CERN and SLC/SLAC have reached a precision,
which goes beyond all former expectations. 
It turns out that we are approaching 
the limits of our theoretical understanding and in order to fully
evaluate the present and future experimental data (LEP1 data still to be
investigated, LHC and a possible $e^+ e^-$ linear collider)
a detailed analysis of 
higher order corrections is necessary. In particular 
at higher energies radiative
corrections become more and more important since they can grow strongly.

Quite a number of groups have started to develop methods for the evaluation
of two-loop diagrams. Analytic results can in general only be obtained for 
specific mass relations, like e.g. all masses zero or 
only one non-zero mass in a diagram. 
Since in the two-loop order for most of the interesting processes the number of
diagrams can become of the order of 1000, it is also necessary to calculate
these diagrams with extremely high precision for the many different masses
occurring in the Standard Model (SM).

  A method which provides particular high precision 
for many kinematical configurations
of interest is that of Taylor expansion in an external 
momentum squared improved by conformal mapping and
the Pad\'e summation technique \cite{Pade}. 
Another method, which is particularly
convenient to program in a formulae manipulating 
language (we use FORM \cite{FORM}), 
is the large mass expansion (LME) \cite{asymptotic}.  
If we have, e.g., diagrams with one large mass like 
a top-quark in a virtual line \cite{ZbbF,Z}, the method is shown to 
be appropriate.
To obtain a satisfying precision, however, it turns out 
to be absolutely necessary 
that higher order terms of the LME are taken into account.
For two-loop propagator-type diagrams such an analysis 
was carried out in \cite{propagator}. In this paper we want
to consider three-point functions.
The particular process we have in mind is $Z \to b\bar{b}$, 
i.e. we investigate vertex diagrams contributing to this 
process which contain a top-quark and compare the results
of the above two methods in order to find out to what precision 
it is possible to
obtain results from the LME. In most of the cases under consideration
the Taylor expansion needs only up to 9 coefficients for a 
very high precision since 
we can confine ourselves here to momenta squared below the lowes threshold.

\section{Large Mass Expansion (LME)}

  Two-loop diagrams as shown in Fig.1
with two different masses on virtual lines, one of which
a top, arise in the process $Z \to b\bar{b}$.   
$W$ and $Z$ are the gauge bosons with masses $M_W$ and $M_Z$, respectively;
$\phi$ is the charged would-be Goldstone boson (we use the Feynman gauge);
$t$ and $b$ are the t- and b-quarks. 
In our analysis we neglect the $b$-quark mass ($m_b=0$). 
Thus we have the following kinematics (see Fig.1):
$p_1^2 = p_2^2 = 0$, $(p_2-p_1)^2 = q^2$ and $q^2 = M_Z^2$
for the on-shell Z. For such type of diagrams analytic results are 
impossible to obtain with presentday technology. 

  Applying the method of the LME, $M_W$ and $M_Z$ are considered
as small. In the framework of this expansion
contributions from additional subgraphs are to be taken into account
together with the Taylor expansion of the initial diagrams w.r.t. external
momenta and light masses. These subgraphs restore the analytic 
properties of the initial diagrams (like logarithmic behaviour).
In other words, the Taylor expansion of the initial diagrams produces
extra infrared singularities which are compensated by singularities of the
additional subgraphs so that only the singularities of the original diagrams 
survive.

For a given scalar
graph $\Gamma$ the expansion in the large mass is given by the formula
\cite{asymptotic}
\begin{equation}
  F_{\Gamma}(q, M ,m, \varepsilon) \stackrel{M \to \infty}{\sim }
  \sum_{\gamma} F_{\Gamma/\gamma}(q,m,\varepsilon) \circ
  T_{q^{\gamma}, m^{\gamma}}
  F_{\gamma}(q^{\gamma}, M ,m^{\gamma}, \varepsilon),
\label{Lama}
\end{equation}
where $\gamma$'s are subgraphs involved in
the LME, $\Gamma/\gamma$ denotes shrinking of $\gamma$ to a
point; $F_{\gamma}$ is the Feynman integrand corresponding to
$\gamma$; $ T_{q_{\gamma}, m_{\gamma}} $ is the Taylor operator
expanding this integrand in small masses $\{ m_{\gamma} \}$ and
external momenta $\{ q_{\gamma} \}$ of the subgraph $\gamma$
; $ \circ$ stands for the convolution of the subgraph expansion
with the integrand $F_{{\Gamma}/{\gamma}}$. The sum goes over all
subgraphs $\gamma$ which (a) contain all lines with large masses, and
(b) are one-particle irreducible w.r.t. light lines.

   At this point it becomes clear what the difference is between the
small-$q^2$ expansion and what is called here the LME:
in the former case we assume all masses large, i.e. 
$q^2 \ll M_W^2,M_Z^2,m_t^2$ while in the latter case only $m_t$ is considered
as large and all other parameters small, i.e. $q^2,M_W^2,M_Z^2 \ll m_t^2$.
In this sense both methods are LM expansions. The technical
advantage of the second method is, however, that only bubbles with one mass occur, 
which can be expressed in terms of $\Gamma$ functions, while in
the case of the small $q^2$ expansion bubbles with different masses are involved,
which are much more difficult to evaluate. Of course also the number and structure
of the subgraphs is different in the two cases.

The structure of the LME of the diagrams under
consideration is given in Fig.2. Bold, thin and
dashed lines correspond to heavy-mass, light-mass, and massless
propagators, respectively. Dotted lines indicate the lines omitted
in the original graph $\Gamma$ to yield the subgraph $\gamma$. 
$\Gamma/\gamma$ (see (\ref{Lama})) then consists out of all the dotted lines
after schrinking $\gamma$ to a point. Thus, in case 0 the last and in case N 
the two last contributions vanish in dimensional regularization.
Finally the LME of the above diagrams has the
following general form:

\begin{equation}
F_{\rm as}^N = \frac{1}{m_t^4}
  \sum_{n=-1}^N \sum_{i,j=-1;i+j=n}^n 
  \left( \frac{M_W^2}{m_t^2} \right)^i \left( \frac{q^2}{m_t^2} \right)^j
  \sum_{k=0}^m A_{i,j,k}(q^2,M_W^2,{\mu}^2) \ln^k \frac{m_t^2}{\mu^2}
\label{series}
\end{equation}
where $m$ is the highest degree of divergence (ultraviolet, infrared, collinear)
in the various contributions to the LME ($m\le$ 3 in the 
cases considered). 
$M_W^2/m_t^2$ and $q^2/m_t^2$ are considered as small parameters. 
$A_{i,j,k}$ are in general complicated functions of the arguments, i.e. they
may contain logarithms and higher polylogarithms 
(see also the explicit examples below). 
(\ref{series}) implies that the difference $F-F^N_{\rm as}$ of the 
integral $F$ and its approximation behaves like 
\begin{equation}
F - F_{\rm as}^N =  O
\left( \frac{
\left( M_W^2 \right)^i \left( q^2 \right)^j \ln^m \frac{m_t^2}{\mu^2}}
{\left(m_t^2 \right)^{N+2}} \right)\,, \qquad \mbox{with } i+j=N\,.
\label{remainder}
\end{equation}

\noindent
One might expect this difference to indicate the order of the 
real error of the asymptotic expansion.
For $q^2$ small this will indeed be the case, for $q^2 \simeq M_Z^2$ the ``small''
parameters are not really small anymore and the validity of the LME
can only be inferred from the comparison with results from other methods, for
which we take here the small $q^2$ expansion. In particular nothing can be said
about a range of convergence of the LME. We can, however, at least
formally apply the Pad\'{e} approximation method to improve the convergence,
the encouraging results of which will be demonstrated below.

\section{Results}

The small momentum expansion of cases 7, 7.1 and 7.2 
(notation due to FST of \cite{Pade})
is described
in detail in Ref. \cite{all}. The additional 
subgraph arising in these cases 
is the same irrespectively of the mass distribution
and is shown in Fig.2 (case 7).   

In the case of the LME two additional subgraphs arise in
each of the cases 7.1 and 7.2. Beyond that these sets of additional
subgraphs are also of quite different structure. Furthermore the $2^{\rm nd}$ 
and $4^{\rm th}$ 
graphs in the row (see Fig.2) produce $1/\varepsilon^3$
terms which cancel, however. Since there are no UV divergences, these must
be mixtures of infrared and collinear ones. 
They determine the highest degree of divergence in these
cases and thus the highest power of the logarithm as discussed in (2). 

Our numerical results for cases 7.1 and 7.2 are presented in Fig.3,
and for $q^2=M_Z^2$ in Table 1. In the figures we show the small 
$q^2$-expansion in comparison with the lowest order approximation
and the sum of terms with N=9, see (2). 
The scale
parameter $\mu=m_t$, i.e. only k=0 contributes in (2). We see that up to 
$q^2=M_Z^2$ the sum of 11 terms agrees quite well with the 
small $q^2$-expansion,
while for higher $q^2$ the agreement very quickly worsens. For this reason we
formally apply also the Pad\'{e} summation technique. With 11
terms in the series, a [5/5]-approximant (long-dashed) can be constructed. It
is seen that indeed this improves the situation considerably 
up to the first threshold
though for a better agreement many more terms in the LME would be
needed. In the small $q^2$-expansion only 9 terms were taken into account, i.e.
a [4/4] approximant is calculated. The high precision of the 
results in this case 
even up to the threshold (in case 7.2) is due to the mapping 
applied in addition 
(see \cite{Pade}). From Table 1 we see that for $q^2=M_Z^2$ 
indeed a rather precise 
result can be obtained with 11 terms from the LME, in particular
if Pad\'e's are applied.

In case 0 an extremely high agreement between the small-$q^2$ expansion and the
LME is obtained (see Table 1). No deviation in a figure like
above would be seen.

The last two cases: A (for asymmetric) and N (for nonplanar) are difficult
from the point of view of the small $q^2$-expansion. While for the planar and
symmetric diagrams a ``standard'' numerical program was written (applied e.g.
in \cite{all}), this program is not without considerable changes applicable for
an asymmetric diagram and not without even more extensions for a nonplanar one.
Thus in these cases just the analytic expressions for the 
first 9 Taylor coefficients 
of the small $q^2$-expansion were produced with FORM in terms 
of bubble integrals and then
numerically evaluated. The problem is that higher coefficients 
get quite lengthy 
in this manner so that
the FORTRAN programs are difficult even to compile (in the multiple precision
version of D. H. Bailey \cite{Bail})!

In case N the lowest threshold is indeed quite high and therefore
the obtained coefficients yield an extremely precise result for the diagram at
$q^2=M_Z^2$. What concerns the LME, the situation is essentially
the same as in cases 7.1 and 7.2 (see Fig.3 and Table 1).

Case A is interesting due to the fact that the lowest threshold is right at
$q^2=M_Z^2$ (in fact the singular contribution at this point should cancel
against some Bremsstrahlung contribution to the $Z\to b\bar b$ decay). Table 2 
demonstrates
that in this case the LME yields results of extremely high
precision, which perfectly agree with the results from 
the small-$q^2$ expansion
at low $q^2$. Near and above the threshold, however, the precision of the small
$q^2$-expansion with a total of 9 coefficients is getting worse, 
but nevertheless
the agreement is still quite surprising. In this case, obviously, the LME
is superior to the small momentum expansion.

   $F^{(0)}$ in Tables 1 and 2 corresponds to the lowest order of the diagrams
(N=-1 or 0, respectively) and $F^{(4)}$ to the order of expansion performed in
Ref. \cite{HSS}.

%
%   TABLES
%

{\scriptsize
\begin{table}[h]
\caption{Values of diagrams at $q^2=90^2,\,
\mu=m_t=180,\, M_W=80$.}

\begin{tabular}{|l|ll|ll|l|l|} \hline
           & $J_{7.1}$  && $J_{7.2}$ && $J_0$ & $J_N$   \\ \hline
$F^{(0)}$  
   & $6.4$       & $-i30.0$
   & $9.2$       & $-i24.5$
   & $0.64$        
   & $-1.5$          \\ 
$F^{(4)}$ 
   & $11.1$       & $-i17.93$ 
   & $9.4$     & $-i44.848$
   & $0.45750$   
   & $-0.86$         \\ 
$F^{(10)}$ 
   & $10.1$       & $-i17.95218$ 
   & $10.122$     & $-i44.84523$
   & $0.457525$   
   & $-0.718$         \\ 
$[5/5]$    
   & $9.996$       & $-i17.9528$   
   & $10.189$      & $-i44.842$ 
   & $0.457523132$  
   & $-0.7031$    \\ 
small-$q$  
   & $9.992668259$  & $-i17.95215366$  
   & $10.193902102$  & $-i44.845273975$
   & $0.4575231327$ 
   & $-0.7026099843$ \\  \hline
\end{tabular}
\label{LME1}
\end{table}
}

\vspace*{-5mm}
{\scriptsize
\begin{table}[h]
\caption{Case A for different $q^2$.}

\begin{tabular}{|l|l|l|ll|ll|} \hline
$q^2/M_Z^2$  & $0.1$  & $0.9$ & $1.1$ && $2.0$ &  \\ \hline
$F^{(0)}$  
   & $-0.324$
   & $-1.36$ 
   & $-0.067$  &  $-i2.17$        
   & $ 0.463$  &  $-i0.408$        \\ 
$F^{(10)}$ 
   & $-0.3431585700$ 
   & $-1.438044241$
   & $-0.089798389$  & $-i2.290275185$
   & $ 0.48623819$   & $-i0.449860818$    \\ 
small-$q$  
   & $-0.3431585700$
   & $-1.4378$              
   & $-0.093$  & $-i2.25$ 
   & $ 0.482$ & $-i0.445$ \\  \hline
\end{tabular}
\label{LME2}
\end{table}
}

{\bf Acknowledgment}\\
\noindent
We want to thank A.~Kotikov and V.~Smirnov for fruitful discussions. 
M.Yu.K. is grateful to the Volkswagen Stiftung and O.V. to BMBF
for financial support.
Both are grateful to the Physics Department of the University of Bielefeld
for the warm hospitality.

\appendix
\section{Analytical Expressions}

  In this section we present the first two terms of the LME 
for the diagrams under consideration. All Feynman integrals
are normalised as follows
\begin{equation}\label{normalization}
   \frac{(4\pi)^{2\varepsilon}}{(16\pi^2)^2 M^4}
   \frac{\Gamma^2(1+\varepsilon)\,\Gamma^2(1-\varepsilon)}
               {\Gamma(1-2\varepsilon)}J
   = \mu^{4\varepsilon}
    \int \frac{d^d k_1}{i(2\pi)^d}
    \frac{d^d k_2}{i(2\pi)^d} \frac{1}{D_1 D_2 D_3 D_4 D_5 D_6}\,,
\end{equation}
where $D_i=(q_i^2-m_i^2)$ are corresponding inverse propagators and
$d=4-2\varepsilon$ is the space-time dimension.
Instead of the large mass $M$, 
the small mass $m$ and the momentum squared, $q^2$, we
introduce the dimensionless parameters
$z=m^2/M^2$ and $s=q^2/m^2$ and write the
results of the LME as a series in $z$ with each
coefficient being $s$-dependent. 
\begin{eqnarray}
  zs J_{7.1} \z=\z
    - \frac{1}{\varepsilon^2}
    - \frac{1}{\varepsilon}(-L_q-L_M+1)
    - \frac12(L_q+L_M)^2 + (L_q+L_M) - 1 \nonumber\\
  \z-\z z \Biggl[
       \Bigl( \frac{1}{\varepsilon^2}-\frac{1}{\varepsilon}L_q
                   +\frac12 L_q^2 \Bigr)
       \Bigl( L_m-L_M+\frac{12+s}{12} \Bigr)  \nonumber\\
  \z\z + \Bigl( \frac{1}{\varepsilon} - L_q \Bigr) 
         \Bigl( \frac12 (L_M^2 - L_m^2)
          -\frac{24+s}{12}L_M  + L_m +\frac{8-3s}{8}\Bigr) 
          + \frac16(L_m^3-L_M^3)\nonumber\\
   \z\z
      -\frac12 L_m^2 + \frac{24+s}{24}L_M^2
      - \frac{s}{2}L_q + L_m + \frac{7s-16}{8}L_M
      + \frac{16-23s}{16} \Biggr] + O(z^2)\,,\\
%%%%%%%%%%%%%%%%%%%%%%%%%%%%%%%%%%%%%%%%%%%%%%%%%%%%%%%%%%%%%%%%%%%%%%%%
  zs J_{7.2} \z=\z
    -\Bigl( \frac{1}{\varepsilon^2} - \frac{1}{\varepsilon}L_q 
               + \frac12 L_q^2 \Bigr) 
    \Bigl( L_M - L_m + I_1(s) - 1\Bigr)  \nonumber\\
   \z\z
    -\Bigl( \frac{1}{\varepsilon} - L_q \Bigr) \Bigl(
      \frac12 L_m^2 - \frac12 L_M^2 + L_M - I_1(s)L_m 
       + I_2(s) -1 \Bigr)
     - \frac16(L_M^3 - L_m^3)                  \nonumber\\
  \z\z 
     + \frac12 L_M^2  - L_M - \frac12 I_1(s)L_m^2 
     + I_2(s) L_m
      - I_3(s) + 1 + O(z) \,,\\ 
%%%%%%%%%%%%%%%%%%%%%%%%%%%%%%%%%%%%%%%%%%%%%%%%%%%%%%%%%%%%%%%%%%%%%%%%
  J_0 \z=\z
        -1 + \zeta(2)
 - z \Biggl[ L_M - L_m +\frac{144+49s}{72}-\frac{4+s}{2}\zeta(2)\Biggr]
     +O(z^2) \,,\\
%%%%%%%%%%%%%%%%%%%%%%%%%%%%%%%%%%%%%%%%%%%%%%%%%%%%%%%%%%%%%%%%%%%%%%%%
  zs J_{A} \z=\z
   +\frac{1}{2\varepsilon} \log(1-s) 
        -\frac12 \Bigl( (L_m+L_M)\log(1-s) 
         + \log^2(1-s) +  {\rm Li}_2(s) \Bigr)  \nonumber\\
   \z-\z z \Biggl[
       -\frac{1}{24\varepsilon} \Bigl( (1+s)\log(1-s) + s\Bigr) 
        - \frac{s}{24}L_m + \frac{s}{8}L_M + \frac{s}{18}  \nonumber\\
   \z\z +\frac{1+s}{24} \Bigl( ( L_m+L_M-1)\log(1-s) 
            + \log^2(1-s) + {\rm Li}_2(s) \Bigr)
          \Biggr] + O(z^2)\,,\\
%%%%%%%%%%%%%%%%%%%%%%%%%%%%%%%%%%%%%%%%%%%%%%%%%%%%%%%%%%%%%%%%%%%%%%%%
  J_{N}  \z = \z
   - \frac{1}{2\varepsilon} - \frac32 + L_M  
   - z \Biggl[ 
   \frac{1}{\varepsilon} \Bigl( L_m - L_M +\frac{72+7s}{48} \Bigl)
      - \frac12 L_m^2 - L_m L_M \nonumber\\
   \z\z  + \frac32 L_M^2 + 3L_m 
        -\frac{144+7s}{24}L_M + \frac{720+181s}{288} + 2\zeta(2) \Biggr]
            + O(z^2)\,,
\end{eqnarray}
where $L_M=\log(M^2/\mu^2)$, $L_m=\log(m^2/\mu^2)$ and
$L_q=\log(-q^2/\mu^2)$ with $\mu$ being an arbitrary
scale parameter.
The functions $I_n(s)$ can be expressed in terms of polylogarithms
${\rm Li}_k$. Here we give only their integral representations
\begin{equation}
  I_n(s) = \frac12\frac{(-)^n}{n!}
    \int_0^1 \frac{\log^n(1-ts/4)}{\sqrt{1-t}} {\rm d}t\,.
\end{equation}

\newpage

%
%  PICTURES
%
%\begin{figure}[h]
\centerline{\vbox{\epsfysize=80mm \epsfbox{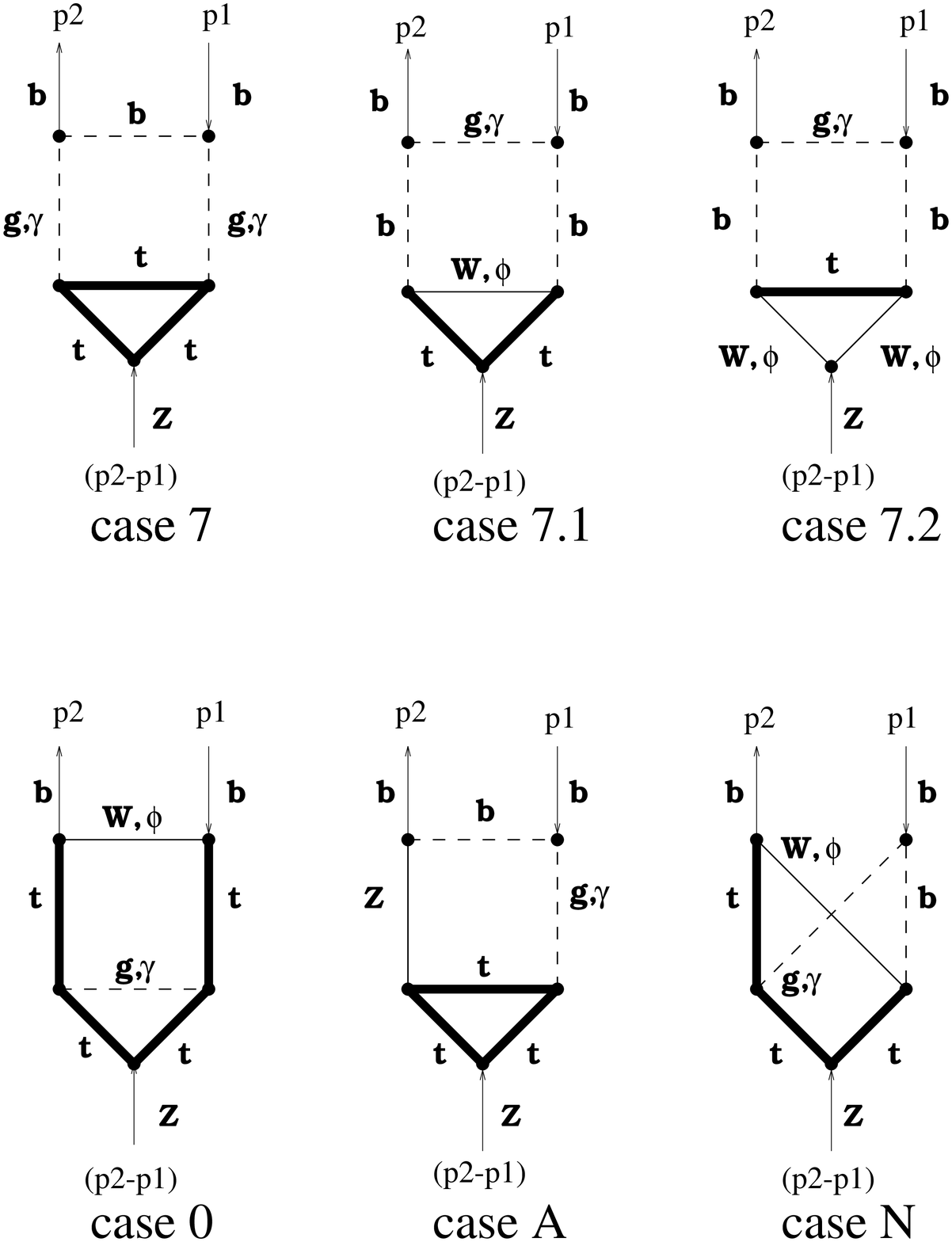}}}
\vspace*{2mm}
\noindent
Figure 1: Two-loop diagrams with two different masses
in internal lines arising in the process $Z  \to b \overline{b}$.
%\caption{\label{diagrams} Two-loop diagrams with two different masses
%in internal lines arising in the process $Z  \to b \overline{b}$ }.
%\end{figure}

\vspace*{10mm}

%\begin{figure}[h]
\centerline{\vbox{\epsfysize=105mm \epsfbox{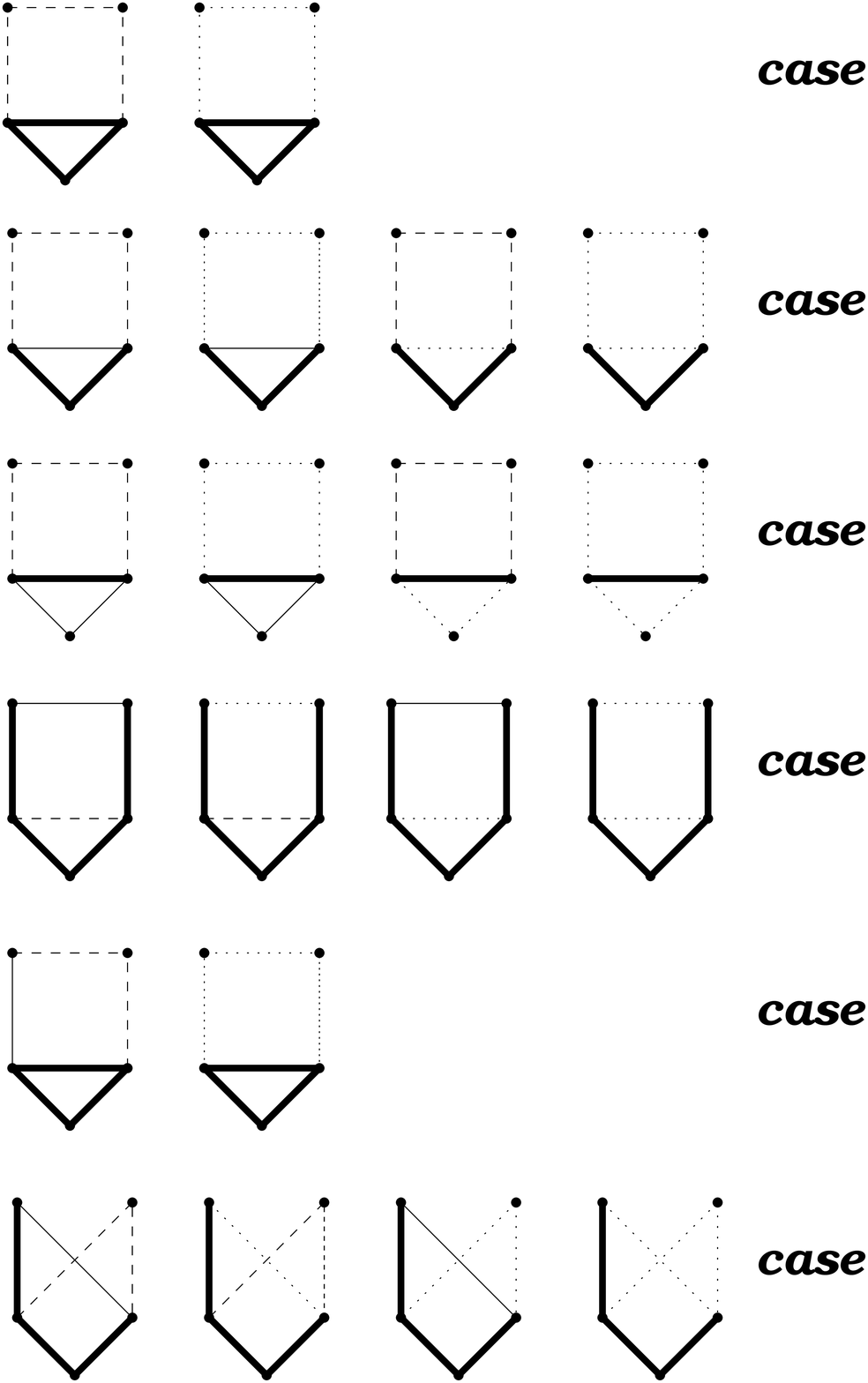}}}
\vspace*{3mm}
\noindent 
Figure 2: The structure of the LME, see explanations in the text.
%\end{figure}
%\caption{\label{expansion} The structure of the LME,
%see explanations in the text.}
%\end{figure}

\newpage

%
%  GRAPHICS'
%
\begin{center}
\vspace*{-3cm}
\vbox{
  \raisebox{6.0cm}{\makebox[0pt]{\hspace*{-2cm}$\mbox{\rm Re}J_{7.1}$}}
  \raisebox{1.5cm}{\makebox[0pt]{\hspace*{18.5cm} $q^2/m_Z^2$}}
  \epsfysize=84mm \epsfbox{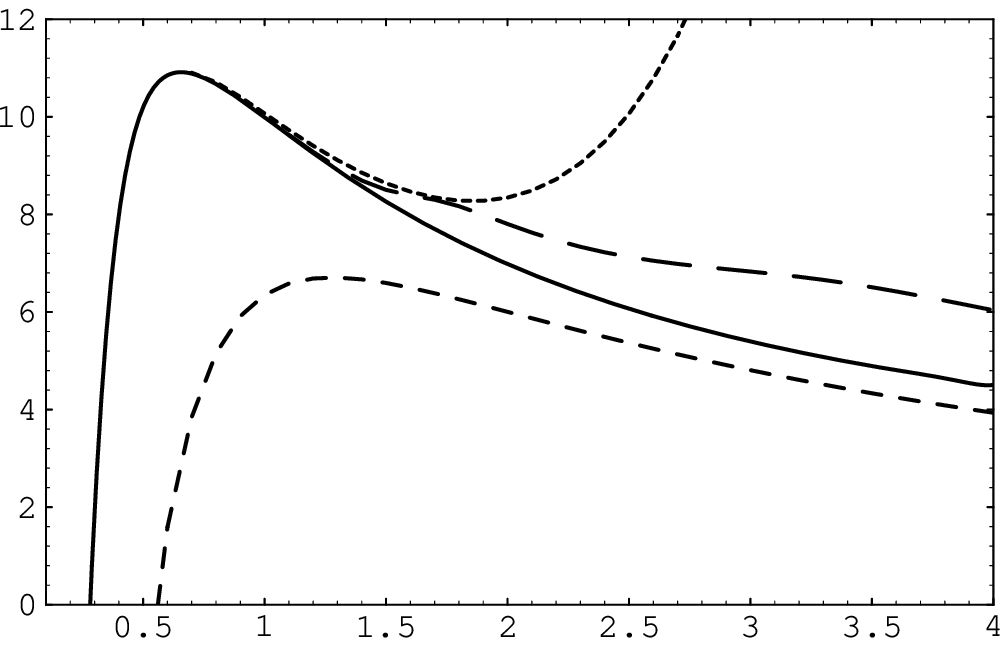}
     }
\vspace*{-2.8cm}
\vbox{
  \raisebox{6.0cm}{\makebox[0pt]{\hspace*{-2cm}$\mbox{\rm Re}J_{7.2}$}}
  \raisebox{1.5cm}{\makebox[0pt]{\hspace*{18.5cm} $q^2/m_Z^2$}}
  \epsfysize=84mm \epsfbox{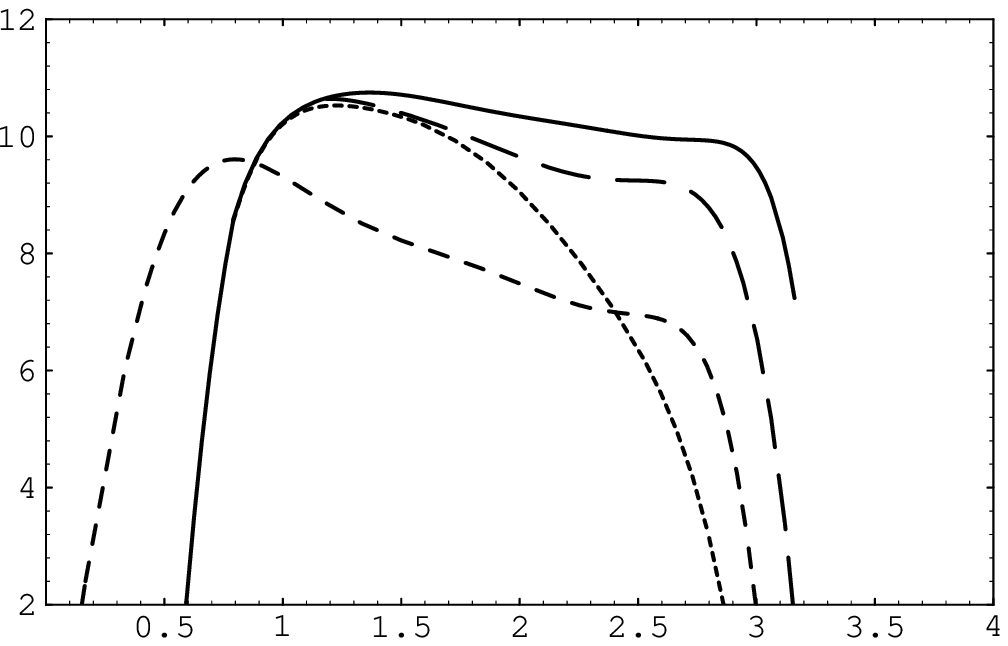}
      }
\vspace*{-3.0cm}
\vbox{
      \raisebox{6.0cm}{\makebox[0pt]{\hspace*{-2cm}$\mbox{\rm Re}J_{N}$}}
      \raisebox{1.8cm}{\makebox[0pt]{\hspace*{18.5cm} $q^2/m_Z^2$}}
      \mbox{\hspace*{-6mm}\epsfysize=90mm \epsfbox{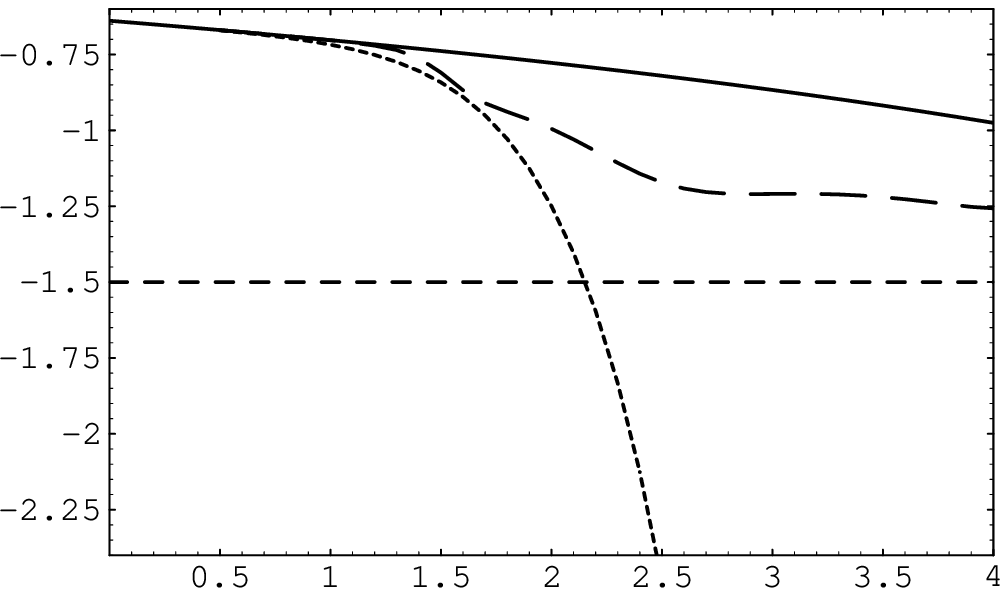}}
      }
\end{center}
\vspace*{-2.0cm}
\noindent
Figure 3: Results of the LME for the real finite parts 
of diagrams 7.1, 7.2 and N with $\mu^2=m_t^2$.
Solid curves represent the small-$q^2$ expansions, 
middle-dashed the leading term of the LME,
short-dashed the sum of 11 terms in the LME,
long-dashed the [5/5] Pad\'e approximant from the LME.

\end{document}